\documentclass{PoS}
\usepackage{bm}

\title{Precision Electroweak measurements at the FCC-ee}

\ShortTitle{Precision Electroweak measurements at the FCC-ee}

\author{\speaker{Mogens Dam}%
  \thanks{On behalf of the FCC Design Study Group.}
  \\
       Niels Bohr Institute, Copenhagen University\\
       E-mail: \email{dam@nbi.dk}}

\abstract{
Because of a luminosity of up to five orders of
magnitude larger than at LEP, electroweak precision measurements at the
FCC-ee -- the Future Circular Collider with electron-positron beams -- would
provide improvements by orders of magnitude over the present status and
constitute a broad search for the existence of new, weakly interacting
particles up to very high energy scales. The FCC-ee will address
centre-of-mass energies ranging from below the Z pole to the
$\mathrm{t\bar{t}}$ threshold and above.
At collision energies around the Z pole, the Z-boson mass and width can be
measured to better than 100 keV each. Asymmetry measurements at the Z pole allow
improvements in the determination of the weak mixing angle by at least a
factor 30 to $\delta\sin^2\theta\mathrm{_W^{eff}}\simeq 6\times 10^{-6}$.
An independent determination of the electromagnetic coupling constant at
the Z energy scale, $\alpha_\mathrm{QED}(m_\mathrm{Z}^2)$, to a relative
precision of $3\times 10^{-5}$ can be obtained via measurement of the
forward-backward asymmetry of lepton pairs at two energy points
$\pm 3.2\,\textrm{GeV}$ away from the Z peak.
At collision energies around the WW threshold, high-statistic
cross section measurements can provide a determination of the W mass to
\mbox{300 keV}.
The key breakthrough advantage of the FCC-ee in these achievements, beside
the large integrated luminosity, is the possibility of a continous,
precise determination of the beam energy by resonant depolarization at the Z
peak and at the WW threshold.
Precise measurements of the hadronic branching fractions of the Z and W
bosons allow for considerably improvements in the
determination of the strong coupling constant down to a precision of
$\delta\alpha_\mathrm{s}(m_\mathrm{Z}^2)\simeq 0.0001$.
An energy scan around the \mbox{350 GeV} $\mathrm{t\bar{t}}$
threshold allows a 10 MeV measurement of the top-quark mass.
}

\FullConference{The European Physical Society Conference on High Energy
  Physics\\
  22-29 July 2015\\
  Vienna, Austria}

\begin{document}

\section{Introduction}

\noindent The Higgs boson discovery~\cite{HiggsATLAS,HiggsCMS} in 2012
at the relatively low mass of 125~GeV has revived the interest in circular
electron-positron colliders that can serve as, in particular, Higgs factories.
The design study~\cite{FCCstudy} of the Future Circular Collider (FCC) in a
$\sim$100-km tunnel in the Geneva area was initiated by CERN in
early 2014. The ultimate goal of the FCC programme is a 100~TeV
proton-proton collider. In the current plans, the first step of the FCC
physics programme would exploit a high-luminosity
$\mathrm{e^+e^-}$ collider, called \mbox{FCC-ee}, able to address
centre-of-mass energies ranging from below the Z pole to the
$\mathrm{t\bar{t}}$ production threshold and above. A first look at the
physics case of the FCC-ee can be found in
Refs.\ \cite{FirstLook,theseBlondel}. The use of
techniques inspired from b-factories, very strong focussing combined with
full-energy top-up injection into separate e$^+$ and e$^-$ storage rings,
allows the FCC-ee to achieve very high instantaneous luminosities of
$8.4 \times 10^{36}\,\mathrm{cm^{-2}\,s^{-1}}$ at the Z pole,
$1.5 \times 10^{36}\,\mathrm{cm^{-2}\,s^{-1}}$ at the WW threshold,
$3.5 \times 10^{35}\,\mathrm{cm^{-2}\,s^{-1}}$ at 240 GeV for HZ production,
and
$8.4 \times 10^{34}\,\mathrm{cm^{-2}\,s^{-1}}$ at the $\mathrm{t\bar{t}}$
threshold~\cite{theseKoratzinos}, where four interaction points have been
assumed.
Compared to linear collider projects, the advantages of the
FCC-ee include the higher luminosity, the
possibility to instrument several interaction points, and the possibility to
measure the centre-of-mass energy with a precision of 100 keV.
The luminosity advantage is about tenfold for the
Higgs-strahlung process $\mathrm{e^+e^- \rightarrow ZH}$ at
$E_\mathrm{CM}=240$~GeV increasing steeply at lower
energies. In summary, the FCC-ee
programme aims at collecting $10^{12-13}$ Z decays in an
energy scan around the Z mass, $2\times 10^8$ W pairs at and above
production threshold, $10^6$ Higgs decays at 240~GeV, and $10^6$
$\mathrm{t\bar{t}}$ pairs at and just above threshold.

In this report, I discuss the
prospects for precision electroweak measurements at the Z pole and at the WW
production threshold. Top quark properties, mass and couplings, can be
measured with high precision at the $\mathrm{t\bar{t}}$
production threshold. Here, only the mass measurement is mentioned; a
dedicated discussion of top-quark couplings can be found in
Ref.\ \cite{topJanot} and in these proceedings~\cite{theseJanot}. Also in
these proceedings can be found discussions of the prospects for the FCC-ee
as a Higgs factory~\cite{theseKlute} and for high precision flavour
physics measurements~\cite{theseMonteil}.

\section{Electroweak precision measurements in the Higgs era}

\noindent Electroweak loops have the remarkable property of being sensitive to
the existence of weakly-coupled particles, even if these cannot be directly
produced or observed in current or future experiments. Historically, electroweak
precision measurements have been instrumental in predicting and determining
free parameters of the Standard Model. Now, with the Higgs boson discovered,
all SM particles are known, and, within the SM, there are no free knobs left
to turn. The precision offered by the FCC-ee on all electroweak observables
will therefore allow potential new physics to be identified, and may be used
towards either constraining or fitting the parameters of
Beyond-Standard-Model theories.

\section{Physics at the Z pole}

\noindent One of the cornerstones of electroweak precision measurements is the
determination of the mass and width of the Z boson. These parameters
were determined at LEP via the line-shape scan to precisions of 2~MeV
each. Key to this
measurement was a precise knowledge of the centre-of-mass energy, which
saturated the systematic uncertainty on the Z mass.
At LEP, a relative precision on the beam energy of $2\times 10^{-5}$ was
reached~\cite{LEPZ} with the technique of resonant
depolarisation~\cite{ELEP}. A detailed discussion of this technique for FCC-ee
can be found in these proceedings~\cite{theseKoratzinos}.

It was found at LEP that the intrinsic precision of each individual beam
energy measurement was 100~keV or better. Since, however, this measurement was
performed only in dedicated polarisation runs, the final uncertainty was more
than tenfold larger because the measurements had to be
extrapolated to the conditions of the physics collision runs.
At FCC-ee, it will be possible to exploit the full precision of the technique
by continuously applying it to a few non-colliding bunches (out of several
tens of thousands), leading to uncertainties better than 100 keV on both the Z
mass and width.

The five orders of magnitude larger statistics than at LEP will
allow also for much improved measurements of other Z-pole observables such as Z
partial widths and numerous asymmetries sensitive to the weak mixing angle.
The benefit of the increased statistics can, in particular, be reaped
for the measurement of ratios, where systematic effects tend to cancel.
Important examples are $R_\ell$, the ratio between the hadronic and leptonic
decay widths, and $R_\mathrm{b}$, the ratio between the b-quark and the total
hadronic widths. For these two parameters, respectively, relative precision
of $5\times 10^{-5}$ and $3 \times 10^{-4}$ are realistic targets. This bears
importance for the determination of the strong coupling constant discussed
below.

At LEP, the determination of the effective weak mixing angle,
$\sin^2\theta\mathrm{_W^{eff}}$, was based on a variety of measurements such
as the leptonic and hadronic forward-backward asymmetries and the $\tau$
polarisation in $\mathrm{Z} \rightarrow \tau\tau$ decays. The single most
precise measurement, however, came from the SLD experiment from the inclusive
left-right beam-polarisation asymmetry, $A_\mathrm{LR}$. Studies are ongoing
to understand whether FCC-ee would be able to operate with longitudinally
polarised beams, and whether, in this case, one would be still able to maintain
the required precision on the beam energy calibration.
Even without longitudinally polarised beams, the increased statistics of
FCC-ee will lead to a sizeable improvement compared to the LEP
measurements, which were largely statistics limited.
Detailed studies of the various asymmetry measurements are under way.
A first study \cite{AfbBlondel} of the forward-backward asymmetry for muon
pairs, $A_\mathrm{FB}^{\mu\mu}$, indicates that this channel alone could
provide an improved precision on $\sin^2\theta\mathrm{_W^{eff}}$ by a factor
50 relative to LEP to $6\times 10^{-6}$, where the dominant systematic
uncertainty contribution from the centre-of-mass energy calibration exceeds
the statistical uncertainty by a factor two.

The very large statistics accumulated at FCC-ee should allow a new range of
searches for very rare phenomena and tests of conservation laws that
remain to be investigated. An example, is the search for sterile, right-handed
partners of neutrinos in Z decays~\cite{sterileBlondel, theseBlondel}.

\section{Physics at the WW production threshold}

\noindent
With $O(10^8)$ W pairs collected at and above the production threshold,
FCC-ee would be able to provide precise measurements of W-boson
properties, such as mass, width, and branching fractions.

For the determination of the W mass and width, operation at a few energy points
within one or two GeV from production threshold is particularly
interesting, since, in this narrow region, the WW cross section is
maximally sensitive to the values of the W mass and width.
Again, to perform precise measurements, the accurate centre-of-mass energy
calibration is necessary.
Transverse beam polarisation builds up naturally in a storage ring by the
Sokolov-Ternov effect~\cite{SokolovTernov}. To maintain polarisation
at a useful level, particles have to avoid depolarising resonances spaced by
440.7~MeV in beam energy. In a storage ring, the beam-energy spread scales
approximately as $\sigma_E \propto E^2/\sqrt{\rho}$, where $\rho$ is the
bending radius. The rapid increase with energy of the beam-energy spread
effectively imposes an upper limit on the energy where a useful polarisation
level can be maintained. At LEP the maximum beam energy at which
polarisation was observed was 60.6~GeV. With the larger bending radius of
FCC-ee, beam polarisation should therefore be available at the WW threshold,
allowing a measurement of the W mass to 300 keV.

\section{The top-quark mass}

\noindent
The FCC-ee will enable precise top-quark studies as well,
with over $10^6$ $\mathrm{t\bar{t}}$ pairs produced at and above
production threshold in a clean experimental environment. An important
achievement will be a sizeable
improvement in the measurement of the top-quark mass, which is
currently determined at hadron colliders to a precision of
about 0.5\%, dominated by the theoretical understanding.
At hadron colliders, the mass is determined from the invariant mass of
the decay products. Because of colour reconnection between final state hadrons
these cannot be unambiguously related to a single top quark, and
hence the method is affected by an inherent uncertainty of the order of
$\Lambda_\mathrm{QCD} \sim 500$~MeV in the theoretical interpretation of the
experimental measurement.

From precise measurements of the $\mathrm{e^+e^- \rightarrow t\bar{t}}$
production cross section in a narrow scan region
around threshold, the top-quark mass can be determined with a statistical
precision of the order of 10~MeV. The measurement benefits from the precisely
defined centre-of-mass energy, which, unlike at linear colliders,
is not affected by strong beamstrahlung effects. The threshold behaviour
of the cross section has a sizeable dependence on the strong coupling constant,
and hence the measurement will profit from the independent
determination of $\alpha_\mathrm{s}$ at the Z and WW running as explained below.
The threshold scan method is characterised by
well-controlled theory uncertainties, likely making it the ultimate
measurement of the top-quark mass.

\section{The strong coupling constant}

\noindent
At LEP, a precise measurement of $\alpha_\mathrm{s}(m_\mathrm{Z}^2)$ was derived
from the Z decay width ratio
$R_\ell = \Gamma_\mathrm{had}/\Gamma_\ell$.
By reinterpreting the LEP mesaurement in the light of
\emph{i)} new N$^3$LO calculations,
\emph{ii)} the improved knowledge of $m_\mathrm{top}$,
and \emph{iii)} the knowledge of $m_\mathrm{Higgs}$,
the uncertainty can be now expressed as
\linebreak
\mbox{$\delta \left(\alpha_\mathrm{s}(m_\mathrm{Z}^2)\right)_\mathrm{LEP} = \pm
0.0038\ \textrm{(exp)} \pm 0.0002\ \textrm{(other)}$}, and it is thus
completely dominated by experimental effects. At FCC-ee, with the
much improved experimental precision on $R_\ell$, a precision on
$\alpha_\mathrm{s}(m_\mathrm{Z}^2)$ of 0.00015 is a reasonable target.

In a similar manner, $\alpha_\mathrm{s}(m_\mathrm{W}^2)$ can be derived from a
measurement of the hadronic branching fraction of the W boson,
$B_\mathrm{had} = (\Gamma_\mathrm{had}/\Gamma_\mathrm{tot})_\mathrm{W}$.
At LEP, with $4\times 10^4$ W-pair events, this quantity was measured with a
relative precision of 0.4\%, not precise enough for an interesting constraint
on $\alpha_\mathrm{s}$. With a factor $O(10^4)$ more data at FCC-ee, an
improvement of up to two orders of magnitude on $B_\mathrm{had}$ can be
foreseen, resulting in a absolute uncertainty on
$\alpha_\mathrm{s}(m_\mathrm{W}^2)$ of $\pm 0.00015$, thus matching the
uncertainty derived from Z decays. From a combination of the Z and W
measurements, a precision of 0.0001 seems within reach for
$\alpha_\mathrm{s}(m_\mathrm{Z}^2)$.

A recent review of the current status and future prospects of precision
$\alpha_\mathrm{s}$ measurements, at FCC-ee and elsewhere, can be found in
Ref.\ \cite{alphaS}.

\section{The Z invisible width and the number of neutrinos}

\noindent
At LEP, the measurement of the Z peak hadronic cross section led
to the determination of the number of active neutrino species,
$N_\nu = 2.984 \pm 0.008$, which can be observed to be two standard deviations
below the SM value of 3. This measurement is of great interest since it
constitutes a direct test of the unitarity of the PMNS matrix -- or of the
existence of right-handed neutrinos, as pointed out in Ref.\ \cite{Jarlskog}.
The experimental conditions at FCC-ee will be adequate to improve the
experimental uncertainty considerably, but, to make the measurement
worthwhile, a commensurate effort would have to be invested in the theoretical
calculation of the small-angle Bhabha-scattering cross section used for the
normalisation. Indeed, the current measurement is limited by the theoretical
understanding of this process.
A desirable goal would be to reduce the uncertainty on $N_\nu$
down to 0.001, but it is not clear that this can be achieved from the Z peak
measurement.

Potentially more promising is the use of radiative return processes at
centre-of-mass energies above the Z pole, \emph{i.e.}\ the process
$\mathrm{e^+e^- \rightarrow Z\gamma}$, leading to a very clean photon-tagged
sample of on-shell Z bosons, with which Z properties can be measured. From the
WW threshold scan alone, 10 million Z$\gamma$ events with
$\mathrm{Z\rightarrow\nu\bar{\nu}}$ and a photon inside the detector acceptance
will be produced per experiment. With in addition the statistics from the 240
and 350 GeV running, the invisible width can be measured with a statistical
precision corresponding to 0.001 on $N_\nu$. Recently, it has been suggested
to search for the direct $s$-channel production of Higgs bosons in a dedicated
high-luminosity run at $E_\mathrm{CM} = 125$~GeV~\cite{theseKlute}. Such
a run would be ideal for the the neutrino counting exercise and could reduce
the statistical uncertainty by about a factor of two relative to the runs at
higher energies.

\section{Direct measurement of
  $\bm{\alpha_\mathrm{\bf QED}}$ at the Z-mass scale}

\noindent The exceptional experimental precision of the Z, W, and top-quark
measurements will require a matching precision on the determination of the
electromagnetic coupling constant at the Z-mass scale relevant for electroweak
interactions, $\alpha_\mathrm{QED}(m_\mathrm{Z}^2)$, in order to exploit the
full potential of the FCC-ee to
unravel signs of new physics.
The current determination of $\alpha_\mathrm{QED}(m_\mathrm{Z}^2)$ is based on the
measurement of the fine structure constant at zero momentum transfer,
$\alpha_\mathrm{QED}(0)$, followed by an extrapolation to the Z-mass
scale. The uncertainty on this extrapolation, involving photon self-energy
diagrams, is dominated by the hadronic part, which is obtained by a dispersion
relation over measured low energy hadronic cross sections in
$\mathrm{e^+e^-}$ annihilations. Current calculations result in a relative
uncertainty on $\alpha_\mathrm{QED}(m_\mathrm{Z}^2)$ of
$1.1 \times 10^{-4}$~\cite{alphaMZ}. An improvement by at least a factor of five
is called for to match the FCC-ee experimental precision.
\begin{figure}
  \begin{center}
    \includegraphics[width=0.56\textwidth]{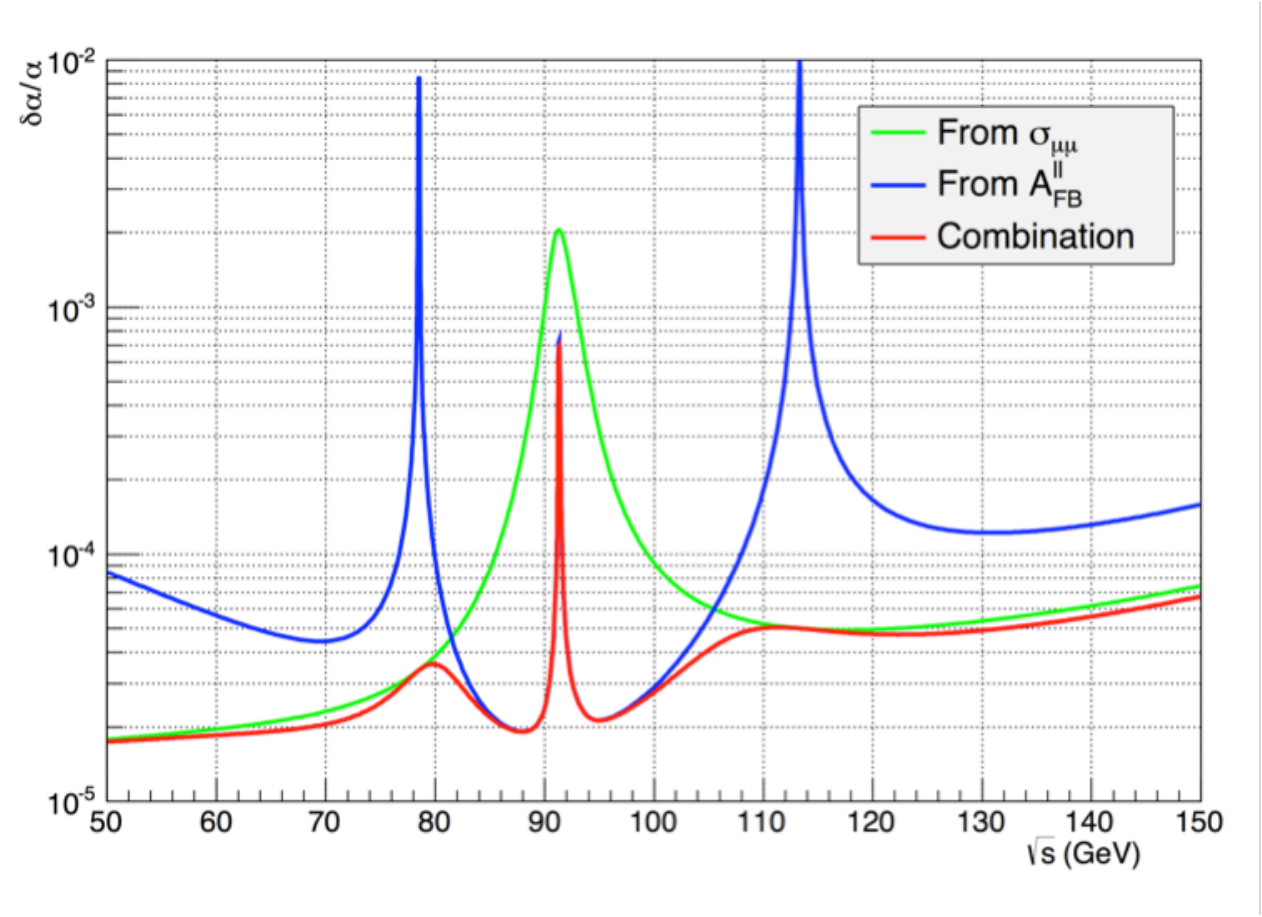}
    \caption{Relative statistical uncertainty on $\alpha_\mathrm{QED}$ versus
      centre-of-mass energy. One year of running at any energy point is
      assumed.}
  \label{fig:alphaQED}
  \end{center}
\end{figure}

It has been recently demonstrated~\cite{alphaQEDJanot1, alphaQEDJanot2} that
the formidable statistics of the FCC-ee could allow a direct measurement of
$\alpha_\mathrm{QED}$ close to the Z-mass scale, eliminating the need for
extrapolation from zero momentum transfer. The study exploits the
$\mathrm{e^+e^- \rightarrow Z}/\gamma^* \rightarrow \ell^+\ell^-$ process. At
the peak of the Z resonance, this process is completely dominated by Z exchange
and has no sensitivity to $\alpha_\mathrm{QED}$. Away from the peak, however,
$\gamma^*$ exchange gradually takes over, and the
sensitivity to $\alpha_\mathrm{QED}$ increases. Two observables have been
investigated: The total production cross section,
$\mathrm{\sigma_{\mu\mu} = \sigma (e^+e^- \rightarrow \mu^+\mu^-)}$ and the
forward-backward asymmetry, $A_\mathrm{FB}^{\ell\ell}$, with $\ell = \mu,
\tau$. Figure~\ref{fig:alphaQED} summaries the study in terms of the
achievable statistical precision on $\alpha_\mathrm{QED}$ after one year of
running at any given energy point in a wide region around the Z pole. Whereas
the statistical uncertainty from the $\sigma_{\mu\mu}$ measurement continues to
decrease away from the Z pole, the uncertainty from the $A_\mathrm{FB}^{\ell\ell}$
measurement shows a more interesting pattern. In fact, the best
relative precision on $\alpha_\mathrm{QED}$ of $2\times 10^{-5}$ appears at
centre-of-mass energies of 87.9 and 94.3 GeV, \emph{i.e.}\ only $\pm 3.2$~GeV
away from the Z peak, energies which could be easily accommodated in the course
of the Z-resonance scan. A comprehensive list of sources of experimental,
parametric, and theoretical systematic uncertainties has been investigated. It
is important to note, that most sources of parametric and theoretical
uncertainties influence the measurement at the 87.9 and 94.3 GeV
energy points in opposite directions, and thus their effects largely
cancel in
the overall measurement. The knowledge of the centre-of-mass energy turns out
to be the dominant systematic uncertainty contribution, smaller than the
statistical uncertainty, though, by a factor two.

\section{Summary}

\begin{figure}
  \begin{center}
    \includegraphics[width=0.57\textwidth]{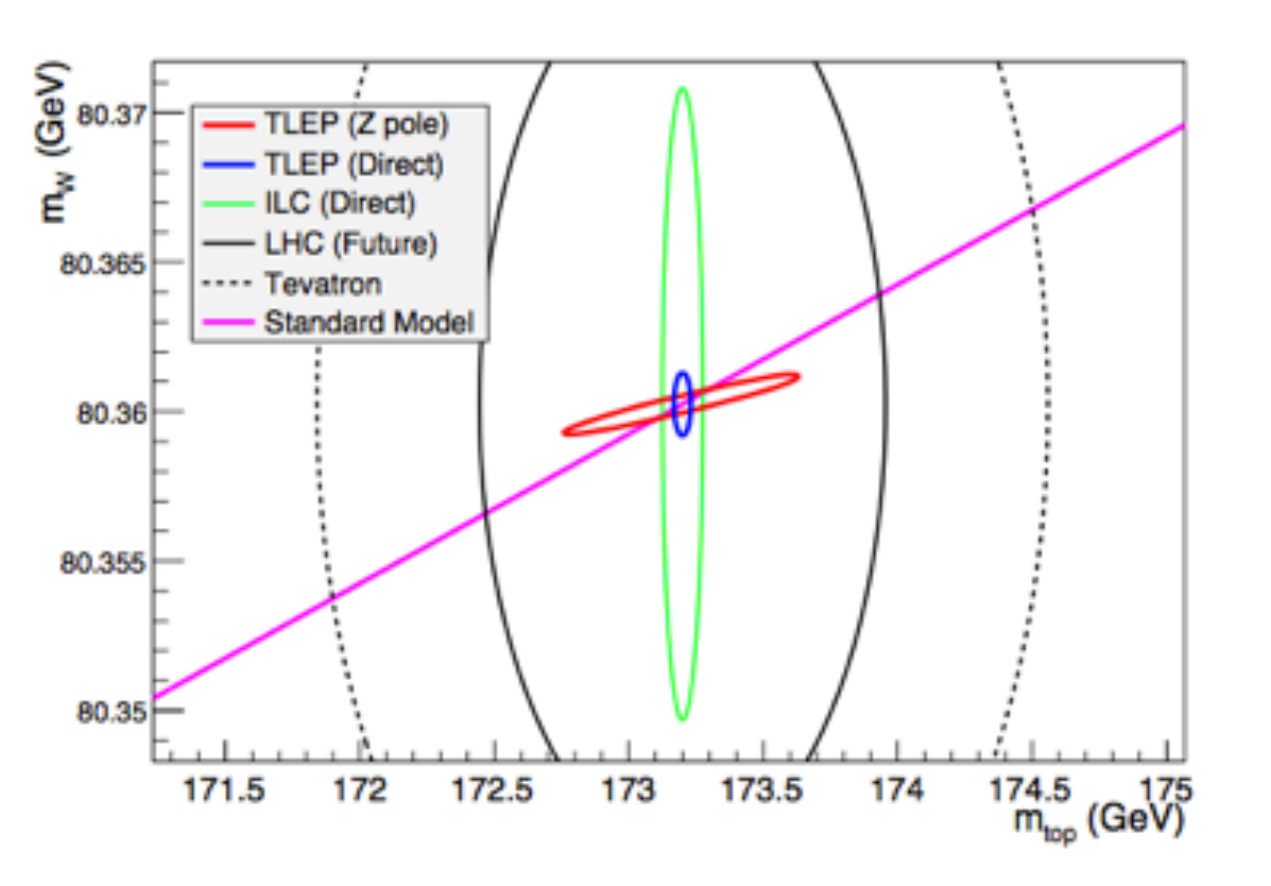}
    \caption{The 68\% c.l.\ contour in the
      $(m_\mathrm{W},\,m_\mathrm{top})$-plane as expected for the FCC-ee
      (indicated as TLEP in the figure) and
      other accelerators. The blue line indicates the expected contour from
      direct W and top-quark mass measurements, while the red line gives the
      expected precision from a fit to the Z-pole observables.}
  \label{fig:contour}
  \end{center}
\end{figure}

\noindent
The proposed FCC-ee, a large, circular $\mathrm{e^+e^-}$ collider
delivering very high luminosities at centre-of-mass energies from 90 to 350 GeV
and above, will allow measurements of electroweak observables at an unrivalled
level of precision. It would provide measurements of the Z-boson
mass and width to better than 100 keV each, the W-boson mass to 300 keV, and
the top-quark mass to 10~MeV.
These and other selected electroweak precision measurements at the FCC-ee are
summarised in Table~\ref{tab:summary},
where the quoted systematic uncertainties are indicative and will be revisited
in the course of the ongoing design study.
\begin{table}
  \begin{center}
  \footnotesize
  \begin{tabular}{|c|c|c|c|c|c|} \hline
    Observable & Measurement & Current precision & FCC-ee stat. &
    Possible syst. & Challenge \\ \hline \hline
    $M_\mathrm{Z}$ [MeV] &
    Lineshape &
    $91187.5 \pm 2.1$ &
    0.005 &
    $<0.1$ &
    QED corr. \\ \hline
    $\Gamma_\mathrm{Z}$ [MeV] &
    Lineshape &
    $2495.2 \pm 2.3$ &
    0.008 &
    $<0.1$ &
    QED corr. \\ \hline
    $R_\ell$ &
    Peak &
    $20.767 \pm 0.025$ &
    0.0001 &
    $<0.001$ &
    Statistics \\ \hline
    $R_\mathrm{b}$ &
    Peak &
    $0.21629 \pm 0.00066$ &
    0.000003 &
    $<0.00006$ &
    $g\rightarrow \mathrm{b\bar{b}}$ \\ \hline
    $N_\nu$ &
    Peak &
    $2.984 \pm 0.008$ &
    0.00004 &
    $0.004$ &
    Lumi meast. \\ \hline
    $A_\mathrm{FB}^{\mu\mu}$ &
    Peak &
    $0.0171 \pm 0.0010$ &
    0.000004 &
    $<0.00001$ &
    $E_\mathrm{beam}$ meast. \\ \hline
    $\alpha_\mathrm{s}(M_\mathrm{Z})$ &
    $R_\ell$ &
    $0.1190 \pm 0.0025$ &
    0.000001 &
    $0.00015$ &
    New Physics \\ \hline
    $1/\alpha_\mathrm{QED}(M_\mathrm{Z})$ &
    $A_\mathrm{FB}^{\mu\mu}$ around peak &
    $128.952 \pm 0.014$ &
    0.004 &
    0.002 &
    EW corr. \\ \hline
    $M_\mathrm{W}$ [MeV] &
    Threshold\ scan &
    $80385 \pm 15$ &
    0.3 &
    $<1$ &
    QED corr. \\ \hline
    $N_\nu$ &
    $\mathrm{e^+e^- \rightarrow \gamma\, Z(inv.)}$ &
    $2.92 \pm 0.05$ &
    0.0008 &
    $<0.001$ &
    ? \\ \hline
    $\alpha_\mathrm{s}(M_\mathrm{W})$ &
    $B_\mathrm{had} = (\Gamma_\mathrm{had}/\Gamma_\mathrm{tot})_\mathrm{W}$ &
    $B_\mathrm{had} = 67.41 \pm 0.27$ &
    0.00018 &
    $0.00015$ &
    CKM Matrix \\ \hline
    $m_\mathrm{top}$ [MeV]&
    Threshold scan &
    $173200 \pm 900$ &
    10 &
    10&
    QCD \\ \hline
  \end{tabular}
\end{center}
  \caption{Selected set of precision measurements at FCC-ee. The systematic
    uncertainties are indicative and will be revisited in the course of the
    design study.}
  \label{tab:summary}
\end{table}
If these goals are achieved, the contour line in the
$(m_\mathrm{top},\,m_\mathrm{W})$-plane could evolve from today to FCC-ee as
indicated in Figure \ref{fig:contour}, where both the results of the direct
mass measurements and the indirect constraints from the precision Z pole
measurements are indicated.
With all of the Standard Model parameters precisely known, the prediction of a
number of observables sensitive to electroweak radiative corrections
become absolute, and any deviation
between measurements would be a demonstration of the existence of new, weakly
interacting particles. With the dramatic increase in precision, sensitivity to
new physics up to energy scales of the order of 100~TeV can be envisioned as
shown in the recent paper~\cite{JEYou}.

\section*{Acknowledgements}

\noindent
I would like to thank Alain Blondel, Patrick Janot, and Roberto Tenchini for
their help in preparing the talk. For his carefull reading of the manuscript,
Patrick Janot deserves another \emph{merci}.

\end{document}